\documentclass{article}
\usepackage{spconf,amsmath,graphicx}
\usepackage{multirow}
\usepackage{booktabs}
\usepackage{xcolor}

\title{Multi-Channel and Multi-Microphone Acoustic Echo Cancellation Using A Deep Learning Based Approach}
\name{Hao Zhang$^1$, DeLiang Wang$^{1,2}$}
\address{
  $^1$Department of Computer Science and Engineering, The Ohio State University, USA\\
  $^2$Center for Cognitive and Brain Sciences, The Ohio State University, USA\\
}

\begin{document}
\ninept
\maketitle
\begin{abstract}

\end{abstract}
Building on the deep learning based acoustic echo cancellation (AEC) in the single-loudspeaker (single-channel) and single-microphone setup, this paper investigates multi-channel AEC (MCAEC) and multi-microphone AEC (MMAEC). We train a deep neural network (DNN) to predict the near-end speech from microphone signals with far-end signals used as additional information. 
We find that the deep learning approach avoids the non-uniqueness problem in traditional MCAEC algorithms. For the AEC setup with multiple microphones, rather than employing AEC for each microphone, a single DNN is trained to achieve echo removal for all microphones. Also, combining deep learning based AEC with deep learning based beamforming further improves the system performance. 
Experimental results show the effectiveness of both bidirectional long short-term memory (BLSTM) and convolutional recurrent network (CRN) based methods for MCAEC and MMAEC.
Furthermore, deep learning based methods are capable of removing echo and noise simultaneously and work well in the presence of nonlinear distortions.

\section{Introduction}
\label{sec:intro}


Acoustic echo cancellation (AEC) is the task of removing undesired echoes that result from the coupling between a loudspeaker and a microphone in a communication system \cite{enzner2014acoustic}. 
Modern hands-free communication devices are usually equipped with multiple microphones and loudspeakers. 
The availability of additional devices also elevates the need for enhanced sound quality and realism, which can hardly be satisfied with single-channel AEC. 
Therefore, it is necessary to design AEC for multiple loudspeakers and/or microphones, which leads to the study of MCAEC and MMAEC.
MCAEC and MMAEC present additional challenges and opportunities compared to single-channel AEC and have received considerable attention recently.


Multi-channel AEC refers to the setup with at least two loudspeakers or channels (stereophonic sound).
Although conceptually similar, MCAEC is fundamentally different from single-channel AEC and a straightforward generalization of single-channel AEC does not result in satisfactory performance because of the non-uniqueness problem \cite{sondhi1995stereophonic}.
This problem is due to the correlation between loudspeaker signals. As a result, the convergence of adaptive technique could be degraded and the echo paths cannot be determined uniquely \cite{sondhi1995stereophonic}.
Many methods have been proposed to circumvent this problem \cite{benesty1995adaptive, shimauchi1997method, schneider2016multichannel, franzen2018efficient}, among which coherence reduction methods are most commonly used. Such methods, however, inevitably degrade sound quality, and a compromise must be made between enhanced convergence and sound quality corruption \cite{sondhi1995stereophonic, luis2019acoustic}.


MMAEC is required for situations in which multiple microphones are present and beamforming techniques are usually combined with AEC for efficient reduction of noise and acoustic echoes.
The most straightforward ways of combining these two processing modules are applying AEC separately for each microphone signal before beamforming or applying a single-microphone AEC to the output of a beamformer \cite{kellermann1997strategies}. 
In general, the former scheme outperforms the latter one \cite{herbordt2001limits}.
Other algorithms employ relative echo transfer functions \cite{reuven2007joint, valero2017multi} or joint optimization strategies \cite{herbordt2004joint, herbordt2000gsaec} to improve the MMAEC performance.
However, efficient combinations of AEC and beamforming are still challenging and many of the strategies exhibit convergence deficiencies \cite{luis2019acoustic}.


Recently, deep learning based methods have been proposed for solving AEC problems and have shown to be effective for echo and noise removal, especially in situations with nonlinear distortions \cite{zhang2018deep, zhang2019deep, carbajal2019joint, sridhar2020icassp}. 
On the basis of the deep learning based single-channel AEC approach, we investigate AEC setups with multiple loudspeakers and microphones. 
The BLSTM based and CRN \cite{tan2018convolutional} based methods are introduced to address MCAEC and MMAEC problems.
Evaluation results show that the
proposed methods effectively remove acoustic echo and background
noise in the presence of nonlinear distortions.

%
The proposed work has four major advantages over traditional methods.
First, although there are multiple acoustic paths in the MCAEC and MMAEC setups, the deep learning based approach can naturally address the problem with model training, rather than employing a separate AEC module for each echo path.
Second, instead of estimating echo paths, deep learning based MCAEC works by directly estimating near-end speech, which intrinsically avoids the non-uniqueness problem.
Third, combining deep learning based AEC and deep learning based beamforming elevates AEC performance remarkably. 
Fourth, deep learning based methods can remove echo and noise simultaneously in the presence of nonlinear distortions.


The remainder of this paper is organized as follows. Section 2 presents the deep learning based approach for MCAEC and MMAEC. Experiments and evaluation results are given in Section 3. Section 4 concludes the paper.

\section{Method Description}

\begin{figure}[!t]
\centering
     \includegraphics[width=0.95\columnwidth,height = 12.5 cm]{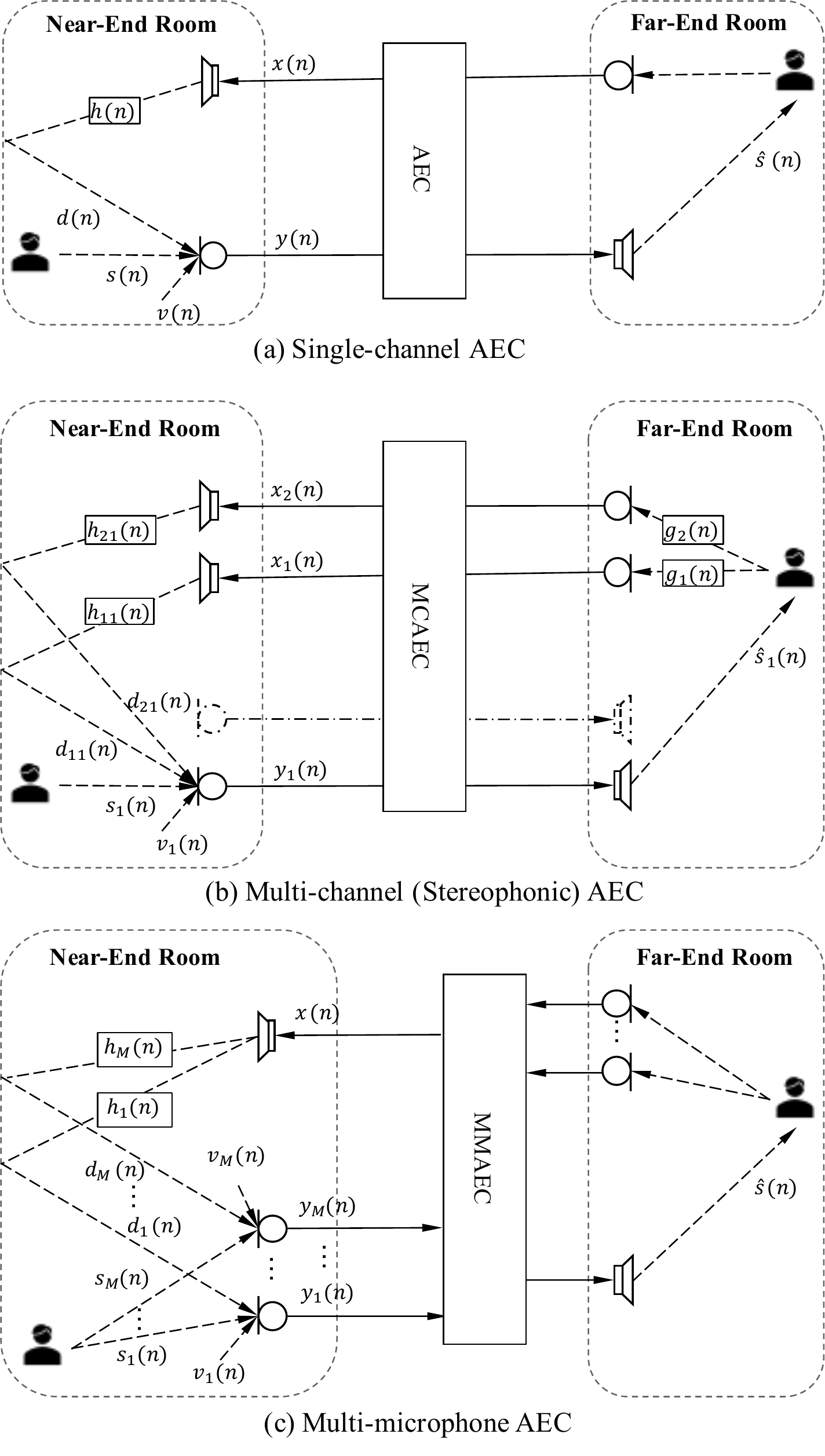}
      \caption{Diagrams of conventional (1) single-channel AEC setup, (2) Multi-channel (Stereophonic) AEC setup, and (c) Multi-microphone AEC setup.}
       \label{fig_AEC}
\end{figure}



\subsection{Deep learning based AEC}

\begin{figure}[!t]
\centering
     \includegraphics[width=0.92\columnwidth, height = 5.5cm]{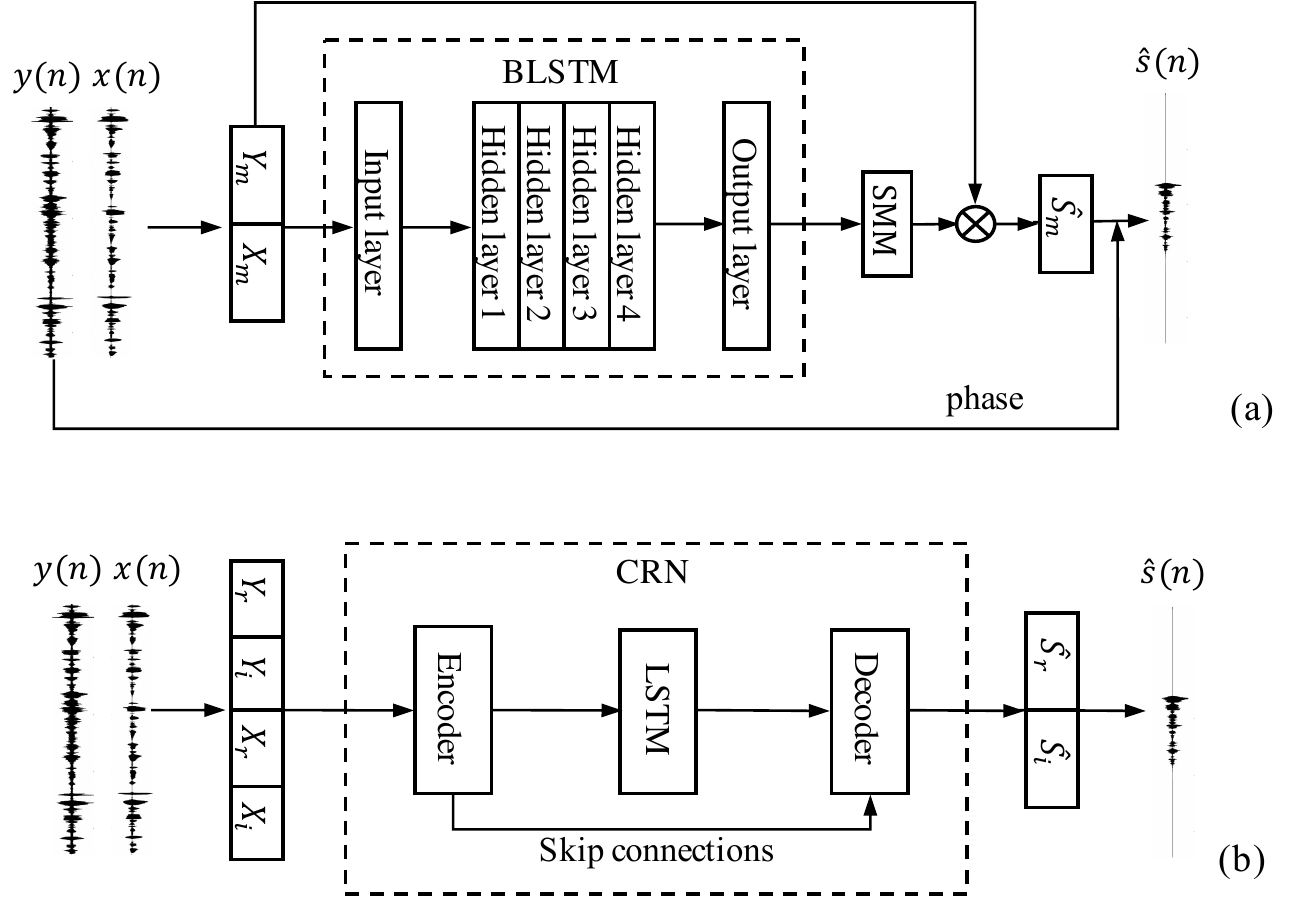}
      \caption{Deep learning for AEC: (a) BLSTM based method, (b) CRN based method. Subscripts ${r}$, ${i}$, and ${m}$ denote real, imaginary and magnitude spectra of signals, respectively.}
       \label{fig_DNN_AEC}
\end{figure}

As is shown in  Fig.~\ref{fig_AEC}(a), the microphone signal $y(n)$ in the single-channel AEC setup is a mixture of echo $d(n)$, near-end speech $s(n)$, and background noise $v(n)$:
\begin{flalign}
\label{equ 1}
y(n) = d(n) + s(n) + v(n)
\end{flalign}
where echo is generated by convolving a loudspeaker signal with a room impulse response (RIR) ($h(n)$).
We formulate AEC as a supervised speech separation problem and the overall approach is to estimate the near-end speech from microphone signal with far-end signal used as additional information. The diagrams of deep learning based methods are shown in Fig.~\ref{fig_DNN_AEC}. 
The input signals, sampled at 16 kHz, are windowed into 20 ms frames with a 10-ms overlap between consecutive frames. Then a 320-point short time Fourier transform (STFT) is applied to
each frame to extract the real, imaginary, and magnitude spectra of signals, which are denoted as $*_{r}$, $*_{i}$, and $*_{m}$, respectively.  

\subsubsection{BLSTM based AEC}
A BLSTM is trained to predict  the spectral magnitude mask (SMM) of near-end speech from the magnitude spectrograms of input signals ($Y_{{m}}$ and $X_{{m}}$), as is shown in Fig.~\ref{fig_DNN_AEC}(a). The SMM is defined as:
\begin{flalign}
SMM(t,f) = \min \{1, |S_{{m}}(t,f)| / |Y_{{m}}(t,f)| \}
\end{flalign}
where $ |S_{{m}}(t,f)|$ and $|Y_{{m}}(t,f)|$ denote the spectral magnitudes of near-end speech and microphone signal within a T-F unit at frame $t$ and frequency $f$, respectively.
Once the model is trained, the estimated spectral magnitude of near-end speech $\hat{S}_m$ is obtained from point-wise multiplication of $Y_{{m}}$ and estimated $SMM$.
Then $\hat{S}_m$, along
with the phase of microphone signal, is sent to the inverse short time
Fourier transform to derive an estimated near-end signal $\hat{s}(n)$.
A detailed description of the BLSTM based method is given in \cite{zhang2018deep}.

\subsubsection{CRN based AEC}

The CRN based method trains the CRN, which is a causal system, for complex spectral mapping \cite{tan2018convolutional}. As is shown in Fig.~\ref{fig_DNN_AEC}(b), it estimates the real and imaginary spectrograms of near-end speech ($\hat{S}_{{r}}$ and $\hat{S}_{{i}}$) from the real and imaginary spectrograms of microphone signal and far-end signal ($Y_{{r}}$, $Y_{{i}}$, $X_{{r}}$, and $X_{{i}}$).
Hence, it is capable of enhancing both magnitude and phase responses simultaneously and $\hat{s}(n)$ resynthesized achieves better speech quality.
A detailed description the CRN architecture is provided in \cite{zhang2019deep}.

\subsection{Deep learning for MCAEC}



Without loss of generality, let us take stereophonic AEC as an example to study deep learning based MCAEC. The signal model is given in Fig.~\ref{fig_AEC}(b) where the stereophonic signals, $x_1(n)$ and $x_2(n)$
are transmitted to loudspeakers and then coupled to one of the microphones. 
The signal picked up by microphone $j$ is composed of two echo signals $d_{1j}(n)$, $d_{2j}(n)$, near-end speech $s_j(n)$, and background noise $v_j(n)$:
\begin{flalign}
\label{equ 1}
y_j(n) = \textstyle \sum_{i=1} ^ 2 d_{ij}(n) +s_j(n) + v_j(n), ~~~~j =1,2.
\end{flalign}
where $d_{ij}(n) = x_i(n)*h_{ij}(n), i =1, 2$, $h_{ij}(n)$ denotes the echo path from loudspeaker $i$ to microphone $j$, and $*$ denotes linear convolution.

Deep learning based MCAEC works by estimating the target $s_j(n)$ given $y_j(n)$, $x_1(n)$, and $x_2(n)$ as inputs.
Specifically, for BLSMT based MCAEC, we use [$Y_{j{m}}$, $X_{1{m}}$, $X_{2{m}}$] as inputs and train the BLSTM to estimate the SMM of $s_j(n)$. For CRN based AEC, we use [$Y_{j{r}}$, $Y_{j{i}}$, $X_{1{r}}$, $X_{1{i}}$, $X_{2{r}}$, $X_{2{i}}$] as inputs and train the network to estimate [${S}_{j{r}}$, ${S}_{j{i}}$]. The training signals are generated by randomly selecting $j$ from $\{1,2\}$, i.e. the model is exposed to signals picked up by the two microphones in MCAEC during training. 
A model trained this way is able to achieve echo removal for both microphones in the system.



\subsection{Deep learning for MMAEC}
\label{sec:MMAEC}



Considering an MMAEC setup with one loudspeaker and $M$ microphones, as is shown in Fig.~\ref{fig_AEC}(b). The signal picked up by microphone $j$ is 
\begin{flalign}
 \textstyle  y_j(n) = d_j(n) + s_j(n) + v_j(n), ~~~~ j =1, 2,\cdots M
\end{flalign}
where $d_j(n) = x(n)*h_j(n)$ is the echo received at microphone $j$.

Different from traditional MMAEC methods that may need to employ a separate AEC module for each microphone in the array, deep learning based MMAEC can be trained to achieve echo removal for all the microphones in the array with a single DNN. 
During training, we use $y_j(n)$ and $x(n)$ as inputs and set the training target to the corresponding near-end speech $s_j(n)$. The training signals are generated by randomly choosing $j$ from $\{1, 2, \cdots, M\}$.

Once the model is trained, the outputs of deep learning based MMAEC could be used for deep learning based minimum variance distortion-less response (MVDR) beamforming \cite{heymann2015blstm}.
Choosing the first microphone in the array as reference microphone, the MVDR beamformer can be constructed as:
\begin{flalign}
 \textstyle  \boldsymbol{\hat{w}}(f) = \frac{\hat{\Phi}_N^{-1}(f)\boldsymbol{\hat{c}}(f)}{\boldsymbol{\hat{c}}(f)^H \hat{\Phi}_N^{-1}(f) \boldsymbol{\hat{c}}(f)}
\end{flalign}
where $(\cdot)^H$ denotes conjugate transpose, $\hat{\Phi}_N(f)$ is the estimated covariance matrix of overall interference (acoustic echo and background noise), $\boldsymbol{\hat{c}}(f)$ is the estimated steering vector, which is estimated as the principal eigenvector of the estimated speech covariance matrix $\hat{\Phi}_s(f)$ \cite{heymann2015blstm,zhang2017speech}.
The covariance matrices of speech and overall interference are estimated from the output of deep learning based MMAEC as
\begin{flalign}
 \textstyle  \hat{\Phi}_S(f) = \frac{1}{T} \sum_t \boldsymbol{\hat{S}}(t,f) \boldsymbol{\hat{S}}^H(t,f)\\
 \textstyle  \hat{\Phi}_N(f) = \frac{1}{T} \sum_t \boldsymbol{\hat{N}}(t,f) \boldsymbol{\hat{N}}^H(t,f)
\end{flalign}
where $\boldsymbol{\hat{S}}(t,f)$ is the STFT representation of estimated speech signals and $\boldsymbol{\hat{N}}(t,f)$ is the estimated overall interference obtained as $\boldsymbol{Y}(t,f) - \boldsymbol{\hat{S}}(t,f)$, $T$ is the total number of frames used in the summation. 

\begin{figure}[!t]
\centering
     \includegraphics[width=0.80\columnwidth, height=2cm]{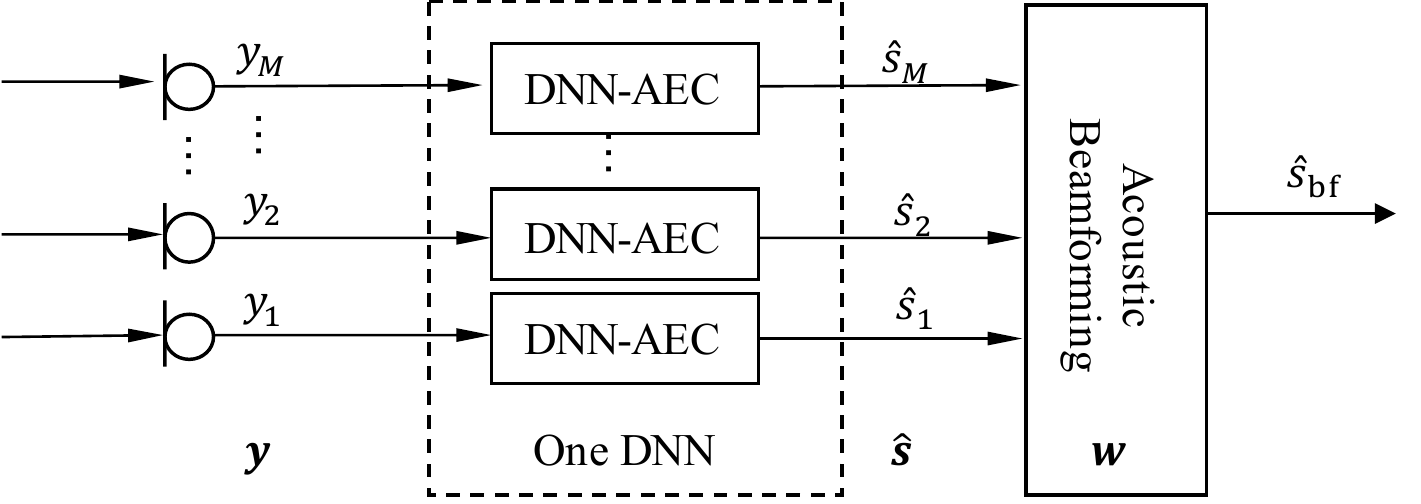}
      \caption{Diagram of combining deep learning based AEC with deep learning based beamforming for further enhancement.}
       \label{fig_AECBF}
\end{figure}

The beamformer is usually applied on microphone signal $\boldsymbol{Y}(t,f)$ and the enhancement results are calculated from
\begin{flalign}
 \textstyle  {Y}_{\text{bf}}(t,f)= \boldsymbol{\hat{w}}^H(f) \boldsymbol{Y}(t,f)
\end{flalign}
Considering that MVDR beamformer performs spatial filtering to maintain signals from the desired direction while suppressing interferences from other directions. A trick we used is to employ the MVDR beamformer as a post-filter for further enhancement. It can be implemented by feeding the output of DNN based MMAEC forward to the deep learning based beamformer with the latter calculated using the same DNN. The overall structure is shown in Fig.~\ref{fig_AECBF}. 
The further enhanced output is obtained using
\begin{flalign}
 \textstyle {\hat{S}}_{\text{bf}}(t,f)= \boldsymbol{\hat{w}}^H(f) \boldsymbol{\hat{S}}(t,f)
\end{flalign}

\section{Experiments}


\subsection{Experiment setting}


The simulation setups for evaluation are designed as follows.
The near-end and far-end speech signals are generated using the TIMIT dataset \cite{lamel1989speech} by following the same way provided in \cite{zhang2019deep}.
Echo signals are generated by convolving far-end signals with RIRs generated using the image method \cite{allen1979image}.
To investigate RIRs generalization, we simulate 20 different rooms of size 
$a\times b \times c$~m (width$\times$length$\times$height) for training mixtures, where $a=[4, 6, 8, 10], b= [5, 7, 9, 11, 13],c=3$. 
For MCAEC setup, the two microphones and the two loudspeakers are positioned at ($a, b + 0.05, c$) m, ($a, b - 0.05, c$) m, ($a, b + 0.6, c+0.5$) m, and ($a, b - 0.6, c+0.5$) m, respectively. 
The near-end speaker is put at 20 random positions in each room with 1 meter apart from the center of the microphones. 
The setup of MMAEC consists of a uniform linear array with four microphones and one loudspeaker. The center of the microphone array is positioned at the center of the room with 4 cm inter-microphone distance. Twenty pairs of positions are simulated randomly for the loudspeaker and the near-end speaker in each room, and the distance from the loudspeaker and the near-end speaker to the center of the array are set to 0.6 m and 1 m, respectively. 
The reverberation time ($T_{60}$) is randomly selected from $\{0.2, 0.3, 0.4, 0.5, 0.6 \}$~s, and the length of RIR is set to 512.
For testing, we simulate three rooms of size $3\times 4 \times 3$~m (Room 1), $5\times 6 \times 3$~m (Room 2), $11\times 14 \times 3$~m (Room3), and set $T_{60}$ to 0.35 s to generate test RIRs for both MCAEC and MMAEC setups.


The most common nonlinear distortion generated by a loudspeaker is the saturation type nonlinearity, which is usually simulated using the scaled error function (SEF) \cite{agerkvist2007modelling, zhang2020deep}:
\begin{flalign}
\label{NMSE} 
f_{\textrm{SEF}}(x) =\int_0^x e^{-\frac{z^2}{2\eta^2}}dz,
\end{flalign}
where $x$ is the input to the loudspeaker, $\eta^2$ represents the strength of nonlinearity. 
The SEF becomes linear as $\eta^2$ tends to infinity and becomes a hard limiter as $\eta^2$ tends to zero. 
To investigate the robustness of the proposed method against nonlinear distortions, four loudspeaker functions are used during the training stage: $\eta^2=0.1$ (severe nonlinearity), $\eta^2=1$ (moderate nonlinearity), $\eta^2=10$ (soft nonlinearity), and $\eta^2=\infty$ (linear). 

Babble noise from NOISEX-92 dataset \cite{varga1993assessment} is used as the background noise and the algorithm proposed in \cite{habets2008generating} is employed to make the noise diffuse.
The diffuse babble noise is then split into two parts, the first $80\%$ of it is used for training and the remaining is used for testing.

We create 20000 training mixtures and 100 test mixtures for both MCAEC and MMAEC setups.
Each training mixture is created by first convolving a loudspeaker signal (generated using randomly selected far-end signal and loudspeaker function) with a randomly chosen training RIR for loudspeaker to generate an echo. 
A randomly chosen near-end utterance is convolved with an RIR for near-end speaker
and then mixed with the echo at a signal-to-echo ratio (SER) randomly chosen from $\{-6, -3, 0, 3, 6\}$ dB. 
Finally, the diffuse babble noise is added to the mixture at a  
signal-to-noise ratio (SNR) randomly chosen from $\{8, 10, 12, 14\}$ dB.
The SER and SNR, which are evaluated during double-talk periods, are defined as:
\begin{flalign}
\textrm{SER} = \textstyle 10\log_{10}\left[\sum_n s^2(n)/ \sum_n d^2(n)\right] 
\end{flalign}
\begin{flalign}
\textrm{SNR} = \textstyle 10\log_{10}\left[\sum_n s^2(n)/ \sum_n v^2(n)\right]
\end{flalign}
Test mixtures are created similarly but using different utterances, noises, RIRs, SERs and SNRs.

The AMSGrad optimizer \cite{reddi2019convergence} and the mean squared error (MSE) cost function are used to train both BLSTM based and CRN based methods. The networks are trained for 30 epochs with a learning rate of 0.001.
The performance of MCAEC and MMAEC is evaluated in terms of utterance-level echo return loss enhancement (ERLE) \cite{enzner2014acoustic}
for single-talk periods and perceptual evaluation of speech quality (PESQ) \cite{rix2001perceptual} for double-talk periods.

\subsection{Performance of MCAEC methods}


We first evaluate the performance of deep learning based MCAEC. The proposed methods are compared with the stereophonic version of joint-optimized normalized least mean square algorithm \cite{paleologu2015overview} equipped with a coherence reduction technique proposed in \cite{djendi2012efficient} (SJONLMS). 
And post-filtering (PF) \cite{ykhlef2014post} is employed to further suppress noises and residual echo (SJONLMS-PF). 
The parameters for these methods are set accordingly to the values given in \cite{paleologu2015overview, djendi2012efficient, ykhlef2014post}.
The comparison results are given in Table~\ref{table_MCAEC}.
In general, the proposed BLSTM based and CRN based MCAEC methods outperform conventional methods and the performance generalizes well to untrained RIRs. BLSTM based method achieves better ERLE results while the CRN based method outperforms all other methods in terms of PESQ. 

\begin{table}[!t]
\centering
\caption{Performance of MCAEC methods in the presence of double-talk, background noise with 3.5 dB SER, 10 dB SNR, $\eta^2 = \infty$ (linear system).}
\label{table_MCAEC}
\resizebox{0.46\textwidth}{!}{%
\begin{tabular}{l | llllll}\specialrule{1.5pt}{1pt}{1pt}
Microphone 1 & \multicolumn{3}{c}{ERLE}                                                                & \multicolumn{3}{c}{PESQ}                                                                \\ \hline
RIRs         & Room1                       & Room2                       & Room3                       & Room1                       & Room2                       & Room3                       \\ \hline
Unprocessed  & -   & -   & -   & 2.12 & 2.13 & 2.17\\ 
SJONLMS      & 7.54 &  7.63 & 7.62 & 2.41 & 2.45 & 2.47 \\
SJONLMS-PF   & 19.04                       & 18.88                       & 18.67                       & 2.45                        & 2.48                        & 2.53                        \\
BLSTM        & \textbf{43.19}              & \textbf{67.17}              & \textbf{67.54}              & 2.57                        & 2.67                        & 2.76                        \\
CRN          & 32.68                       & 35.34                       & 41.31                     & \textbf{2.62}               & \textbf{2.83}               & \textbf{2.98}               \\ \specialrule{1.5pt}{1pt}{1pt} \specialrule{1.5pt}{1pt}{1pt}

Microphone 2 & \multicolumn{3}{c}{ERLE}                                                                & \multicolumn{3}{c}{PESQ}                                                                \\ \hline
RIRs         & Room1                       & Room2                       & Room3                       & Room1                       & Room2                       & Room3                       \\ \hline
Unprocessed  & -   & -    & -    & 2.11 & 2.14 & 2.18 \\
SJONLMS      & 7.51                        & 7.76                        & 7.63                        & 2.41                        & 2.45                        & 2.45                        \\
SJONLMS-PF   & 18.98                       & 18.89                       & 18.60                       & 2.45                        & 2.50                        & 2.50                        \\
BLSTM        & \textbf{49.12}              & \textbf{67.47}              & \textbf{67.67}              & 2.55                        & 2.68                        & 2.76                        \\
CRN          & 31.25                       & 36.61                       & 41.57                       & \textbf{2.60}               & \textbf{2.85}               & \textbf{2.99}           \\ \specialrule{1.5pt}{1pt}{1pt}   
\end{tabular}}
\end{table}

\subsection{Performance of MMAEC methods}


This part studies the performance of deep learning based MMAEC. We employ single-channel JONLMS \cite{paleologu2015overview} for each microphone in the array as a baseline and then combine the outputs with the ideal MVDR beamformer (JONLMS-IBF). The ideal MVDR beamformer (IBF) is calculated by substituting the true speech and interference components of the microphone signal ($ \boldsymbol{{S}}(t,f)$ and $ \boldsymbol{{N}}(t,f)$) into (6), (7), and (5). Therefore, it can be regarded as a stronger baseline compared to other MVDR beamformers. 
Three results are provided for each deep learning based method, in which $\hat{s}$ is the output of the reference microphone, $y_{\text{bf}}$ and $\hat{s}_{\text{bf}}$ are, respectively, the time-domain beamformed microphone signal and beamformed enhanced signal introduced in Section~\ref{sec:MMAEC}. The comparison results are given in Table~\ref{table_MMAEC}. As can be seen from the table, deep learning based methods outperform traditional MMAEC methods in terms of ERLE and PESQ. 
Single-channel outputs of deep learning based methods ($\hat{s}$) are good enough for echo and noise removal while combining deep learning based beamformer as a post-filter ($\hat{s}_{\text{bf}}$) further improves the overall performance in most of the cases.

\begin{table}[!t]
\centering
\caption{Performance of MMAEC methods in the presence of double-talk, background noise with 3.5 dB SER, 10 dB SNR, $\eta^2 = \infty$ (linear system).}
\label{table_MMAEC}
\resizebox{0.46\textwidth}{!}{%
\begin{tabular}{ll | llllll} \specialrule{1.5pt}{1pt}{1pt}
\multicolumn{2}{l|}{}           & \multicolumn{3}{l}{ERLE}                                                                         & \multicolumn{3}{l}{PESQ}                                                                                  \\  \hline
\multicolumn{2}{l|}{RIRs}           & Room1                          & Room2                          & Room3                          & Room1                             & Room2                             & Room3                             \\  \hline
\multicolumn{2}{l|}{Unprocessed}           & -& - &-&2.04& 2.09 & 2.10 \\
\multicolumn{2}{l|}{JONLMS}           & 6.94    & 6.95   & 6.93   & 2.43   & 2.45       & 2.48       \\
\multicolumn{2}{l|}{JONLMS-IBF}           & 17.61                          & 16.76                          & 15.52                          & 2.70         & 2.63                              & 2.66                              \\  \hline
                        & $\hat{s}$    & \textbf{62.03}                 & 65.65                & \textbf{66.84}                 & 2.57                              & 2.71                              & 2.74                              \\
                        & $y_{\text{bf}}$ & 2.65                           & 5.15                           & 2.06                            & 2.17                              & 2.39                              & 2.20                              \\
\multirow{-3}{*}{BLSTM} & $\hat{s}_{\text{bf}}$ & 60.34                          & \textbf{67.22}                 & {66.11}                 & 2.68          & 2.87                              & 2.76                              \\  \hline
                        & $\hat{s}$    & 25.92                          & 32.94                          & 33.99                          & 2.66                     & 2.89                    & \textbf{2.94} \\ 
                        & $y_{\text{bf}}$ & 2.77                           & 5.48                           & 2.24                           & 2.18                              & 2.41                              & 2.21                              \\
\multirow{-3}{*}{CRN}   & $\hat{s}_{\text{bf}}$ & 27.57                          & 36.92                          & 34.11                          & \textbf{2.75}                     & \textbf{2.98}                     & 2.89        \\  \specialrule{1.5pt}{1pt}{1pt}
\end{tabular}}
\end{table}

\subsection{Performance in situations with nonlinear distortions}


This part tests the performance of BLSTM based and CRN based MCAEC and MMAEC in situations with nonlinear distortions introduced by loudspeaker. The results are shown in Table~\ref{NL_MMAEC}. Comparing the results with those without nonlinear distortions shows that the deep learning based methods can be trained to handle cases with and without nonlinear distortions and the performance generalizes well to untrained nonlinearity ($\eta^2=0.5$).

\begin{table}[!t]
\centering
\caption{Performance of deep learning based MCAEC and MMAEC in the presence of double-talk, background noise and nonlinear distortions with Room2, 3.5 dB SER, 10 dB SNR, $\eta^2 = 0.1$, $\eta^2 = 0.5$.}
\label{NL_MMAEC}
\resizebox{0.46\textwidth}{!}{%
\begin{tabular}{l| ll | llll} \specialrule{1.5pt}{1pt}{1pt}
\multicolumn{3}{l|}{}           & \multicolumn{2}{l}{ERLE}                                                                         & \multicolumn{2}{l}{PESQ}                                                                                  \\  \hline
\multicolumn{3}{c|}{Nonlinearity}           & $\eta^2 = 0.1$                          & $\eta^2 = 0.5$                                       & $\eta^2 = 0.1$                             & $\eta^2 = 0.5$                                                  \\  \hline

\multirow{3}{*}{{\begin{tabular}[c]{@{}l@{}}MCAEC\\ (Room2 \\Microphone 1)\end{tabular}}} & \multicolumn{2}{l|}{Unprocessed}    & -   & -   & 2.11 & 2.13\\ 
& \multicolumn{2}{l|}{BLSTM}          & {67.32}              & {67.76}                        & 2.67                       & 2.67                                             \\
& \multicolumn{2}{l|}{CRN}            & 34.86                      & 34.72                                         & {2.82}               & {2.83}                         \\ \specialrule{1.5pt}{1pt}{1pt}

\multirow{7}{*}{{\begin{tabular}[c]{@{}l@{}}MMAEC\\ (Room2) \end{tabular}}} & \multicolumn{2}{l|}{Unprocessed}           & -& - &2.08& 2.08 \\ \cline{2-7} 
                    &    & $\hat{s}$    &  65.81                & 64.93                         & 2.70                              & 2.70                                                  \\
              &          & $y_{\text{bf}}$ & 5.11                          & 5.13                                               & 2.38                              & 2.38                                                    \\
& \multirow{-3}{*}{BLSTM} & $\hat{s}_{\text{bf}}$ & 67.18                          & 66.55                            & 2.87          & 2.86                                               \\  \cline{2-7} 
                 &       & $\hat{s}$    & 33.15                       & 33.05                                           & 2.89                    & 2.88           \\ 
              &          & $y_{\text{bf}}$ & 5.49                          & 5.46                                           & 2.40                             & 2.40                                                     \\
& \multirow{-3}{*}{CRN}   & $\hat{s}_{\text{bf}}$ & 36.84                         & 36.83                                             & 2.99                     & 2.99                           \\  \specialrule{1.5pt}{1pt}{1pt}
\end{tabular}}
\end{table}

\section{Conclusion}

We have proposed a deep learning approach to MCAEC and MMAEC. Our approach overcomes the limitations of traditional methods and produces remarkable performance in terms of ERLE and PESQ.
Evaluation results show the effectiveness of both BLSTM and CRN based methods for removing echo and noise in cases with and without nonlinear distortions, and the performance generalizes well to untrained RIRs.
Moreover, the proposed methods can be extended to handle a general AEC setup with an arbitrary number of microphones and an arbitrary number of loudspeakers, which will be demonstrated in future research.

\section{Acknowledgements}
This research was supported in part by an NIDCD grant (R01 DC012048) and the Ohio Supercomputer Center.

\vfill\pagebreak

\bibliographystyle{IEEEbib}
\bibliography{Zhang_2019}

\end{document}